\documentclass[aps,twocolumn,showpacs,nofootinbib,superscriptaddress]{revtex4-1}

\usepackage[utf8]{inputenc}
\usepackage{amssymb}
\usepackage{amsmath}
\usepackage{amsfonts}
\usepackage{dsfont}
\usepackage[usenames,dvipsnames]{xcolor}
\usepackage[pdftex]{graphicx}
\usepackage[pdftex,plainpages=false,colorlinks=true,linkcolor=Red, citecolor=blue, urlcolor=blue]{hyperref}
\usepackage{bm}

\usepackage{txfonts}

\renewcommand{\vec}[1]{\bm{\mathrm{#1}}}

\begin{document}

\title{Intrinsic Cluster-Shaped Density Waves in Cellular Dynamical Mean-Field Theory}

\author{S. \surname{Verret}}
\email[Corresponding author: ]{simon.verret@usherbrooke.ca}
\thanks{additional affiliation: Mila, Universit\'e de Montr\'eal, Qu\'ebec, Canada H2S 3H1}
\affiliation{D\'epartement de physique and RQMP, Universit\'e de Sherbrooke, Qu\'ebec, Canada J1K 2R1}
\affiliation{Institut quantique, Universit\'e de Sherbrooke, Qu\'ebec, Canada J1K 2R1}
\author{J. \surname{Roy}}
\thanks{Current affiliation: Department of Physics, University of Toronto, 60 St. George Street, Toronto, Ontario, Canada M5S1A7}
\affiliation{D\'epartement de physique and RQMP, Universit\'e de Sherbrooke, Qu\'ebec, Canada J1K 2R1}
\affiliation{UM-DAE Centre for Excellence in Basic Sciences, Santa Cruz(E), Mumbai 400098, India}
\author{A. \surname{Foley}}
\affiliation{D\'epartement de physique and RQMP, Universit\'e de Sherbrooke, Qu\'ebec, Canada J1K 2R1}
\affiliation{Institut quantique, Universit\'e de Sherbrooke, Qu\'ebec, Canada J1K 2R1}
\author{M. \surname{Charlebois}}
\thanks{Current affiliation:
Center for Computational Quantum Physics, Flatiron institute, Simons Foundation, 162 5th Ave., New York, 10010 NY, USA}
\affiliation{D\'epartement de physique and RQMP, Universit\'e de Sherbrooke, Qu\'ebec, Canada J1K 2R1}
\affiliation{Institut quantique, Universit\'e de Sherbrooke, Qu\'ebec, Canada J1K 2R1}
\author{D. \surname{Sénéchal}}
\affiliation{D\'epartement de physique and RQMP, Universit\'e de Sherbrooke, Qu\'ebec, Canada J1K 2R1}
\affiliation{Institut quantique, Universit\'e de Sherbrooke, Qu\'ebec, Canada J1K 2R1}
\author{A.-M. S. \surname{Tremblay}}
\affiliation{D\'epartement de physique and RQMP, Universit\'e de Sherbrooke, Qu\'ebec, Canada J1K 2R1}
\affiliation{Institut quantique, Universit\'e de Sherbrooke, Qu\'ebec, Canada J1K 2R1}
\affiliation{Canadian Institute for Advanced Research, Toronto, Ontario, Canada M5G 1Z8}
\date{\today}
\begin{abstract}
It is well known that cellular dynamical mean-field theory (CDMFT) leads to the artificial breaking of translation invariance. In spite of this, it is one of the most successful methods to treat strongly correlated electrons systems. 
Here, we investigate in more detail \emph{how} this broken translation invariance manifests itself. This allows to disentangle artificial broken translation invariance effects from the genuine strongly correlated effects captured by CDMFT.
We report artificial density waves taking the shape of the cluster---cluster density waves---in all our zero temperature CDMFT solutions, including pair density waves in the superconducting state. 
We discuss the limitations of periodization regarding this phenomenon, and we present mean-field density-wave models that reproduce CDMFT results
at low energy in the superconducting state. We then discuss how these artificial density waves help the agreement of CDMFT with high temperature superconducting cuprates regarding the low-energy spectrum, in particular for subgap structures observed in tunnelling microscopy. We relate these subgap structures to nodal and anti-nodal gaps in our results, similar to those observed in photoemission experiments. This fortuitous agreement suggests that spatial inhomogeneity may be a key ingredient to explain some features of the low-energy underdoped spectrum of cuprates with strongly correlated methods.
This work deepens our understanding of CDMFT and clearly identifies signatures of broken translation invariance in the presence of strong correlations.
\end{abstract}

\maketitle

\section{Introduction}

Broken translation invariance has a long history in cuprate high temperature superconductors~\cite{comin_resonant_2016, julien_magnetic_2015}. In addition to the antiferromagnetic order found at zero doping, broken translation invariance was observed as stripe order with neutron scattering~\cite{tranquada_evidence_1995} and as charge order with scanning tunnelling microscopy (STM)~\cite{hoffman_four_2002,kohsaka_intrinsic_2007}. Fermi surface reconstruction caused by broken translation invariance was also deduced from Quantum oscillations and transport measurements~\cite{leboeuf_electron_2007,ramshaw_angle_2011,leboeuf_lifshitz_2011} before charge order was, finally, unambiguously detected in the bulk of cuprates with nuclear magnetic resonance~\cite{wu_magnetic-field-induced_2011,wu_incipient_2015} and resonant X-rays scattering~\cite{ghiringhelli_long-range_2012}. It has since become a landmark of the low-temperature, underdoped region of the phase diagram of cuprates. The charge modulation has a period of three to four unit cells and it is short-ranged and unidirectional, although often appearing as a checkerboard order because of overlapping domains~\cite{hamidian_atomic-scale_2016,comin_broken_2015, comin_resonant_2016}. In the CuO$_2$ planes, the charge is modulated on the oxygen atoms (O$_x$, O$_y$), with a $d$-wave form factor, i.e., maximal on O$_x$ when minimal on O$_y$ and vice-versa~\cite{fujita_direct_2014,comin_symmetry_2015,hamidian_atomic-scale_2016}. Aspiring theories of cuprates must now address such low energy broken translation invariance to be complete.

The regime where translation invariance is broken in cuprates---intermediate interaction strength, low-temperature, and low-doping---is very difficult to address theoretically~\cite{simons_collaboration_on_the_many-electron_problem_solutions_2015}.
Outside this regime, however, strongly correlated electron methods capture most of the phenomenology of cuprates~\cite{tremblay_pseudogap_2006}. In particular, cellular dynamical mean-field theory (CDMFT)~\cite{kotliar_cellular_2001, maier_quantum_2005} predicts a pseudogap in the Hubbard model of cuprates without any need for broken symmetry, due to strong electronic correlations~\cite{kyung_pseudogap_2006,stanescu_fermi_2006,civelli_doping-driven_2009,civelli_nodal-antinodal_2008,gull_momentum-space_2010,Reymbaut:2019}.
In this framework, the decrease in magnetic susceptibility found at a temperature $T^*$ signals the opening of the pseudogap. It is driven by short-range spin correlations in a doped Mott insulator ~\cite{sordi_finite_2010,sordi_mott_2011,fratino_organizing_2016}, a result also found for the three-band Emery model~\cite{fratino_pseudogap_2016}. CDMFT also correctly captures the phase diagram of cuprates for antiferromagnetism and superconductivity~\cite{capone_competition_2006,kancharla_anomalous_2008, foley_coexistence_2019}.
The main unresolved issue is the presence of many unexplained broken symmetries~\cite{fauque_magnetic_2006,Mook:2008,kapitulnik_polar_2009,Shekhter:2013,Lubashevsky:2014,cyr-choiniere_two_2015,Mangin-Thro:2015,zhao_global_2016,Zhao:2017} that appear at a temperature $T^*_{phase}$ which is smaller than $T^*$ in the hole-underdoped compounds. Broken translation invariance is one of these, notably for the aforementionned stripe and charge orders, but also for pair density waves, for which evidence was recently provided by STM~\cite{hamidian_detection_2016,edkins_magnetic-field_2018}.

Whether density waves could be the missing ingredient to improve CDMFT results in the low-temperature low-doping regime proves to be a very challenging question.
Cluster methods such as CDMFT are badly suited for studying density waves beyond simple N\'eel aniferromagnetism because their treatment requires an enlarged unit cell, and cluster methods already rely on an enlarged unit cell: the cluster itself. Therefore, these methods are often restricted to specific wavelengths of density waves, determined by the size of the cluster, or they are biased towards these wavelengths. In this respect, CDMFT has an important well-known problem: it already intrinsically breaks translation invariance.\footnote{Not all cluster extensions of dynamical mean-field theory (DMFT) break translation invariance. The dynamical cluster approximation (DCA)~\cite{hettler_nonlocal_1998,maier_quantum_2005} enforces translation invariance, at the cost of piecewise constant self-energy in reciprocal space, and the periodized-CDMFT (PCDMFT)~\cite{biroli_cluster_2004} attempts to correct CDMFT. We discuss these methods in sections~\ref{sec_translation} and~\ref{sec_periodization}} Nevertheless, it is possible to study density waves with wavelengths longer than the cluster by using multi-cluster methods like I-CDMFT~\cite{charlebois_impurity-induced_2015,faye_interplay_2017-1} to remove some of the bias~\cite{faye_interplay_2017-1}.

The main focus of the present paper is this intrinsic translation invariance breaking of CDMFT. 
We present results showing that, at low energy, CDMFT breaks translation invariance with the same signatures as those of density waves. To show this, we define order parameters for density waves that emulate our cluster tiling, and we show that these order parameters are non-zero in all our CDMFT solutions. Second, we construct a mean-field model to study how $d$-wave superconductivity interplays with these density-waves orders and we show that this non-interacting model reproduces the low energy features of our superconducting CDMFT solution.
These results demonstrate that artificial cluster-shaped density waves are an intrinsic component of CDMFT results.

With the acknowledged presence of those density waves, we revisit the low-energy single-particle spectral weight of the superconducting state found in CDMFT. We find that broken translation invariance is responsible for two important features of the superconducting gap: subgap structures in the density of states, and a differentiation between nodal and anti-nodal gaps in the spectral weight. We then explain that these two features roughly agree with homologous observations in cuprates: sub-gap structures seen in STM~\cite{mcelroy_coincidence_2005, bruer_revisiting_2016} and the nodal and anti-nodal gaps seen in ARPES~\cite{hashimoto_energy_2014}. Of course, this agreement is fortuitous---it is caused by artificial density waves--- and thus requires a nuanced discussion.

Thus, this paper provides the order parameters that measure the intrinsic broken translation invariance of CDMFT, explore in details the consequences of this broken translation invariance, and how these consequences relate to experimental results in cuprates. This allows to disentangle the effects of broken translation invariance from genuine strongly correlated physics both in cuprates and in CDMFT: we conclude by suggesting that the subgap structures and the nodal and antinodal dichotomy of the superconducting gap might be indirect consequences of inhomogeneity rather than direct consequences of strong correlations.

Of course, since the density waves found in CDMFT solutions are an artefact, one may want to remove them. Although this paper does not offer a way to do so, we discuss many important points regarding this endeavour. 
First, we show that using a large Lorentzian broadening to plot the spectral weight de-emphasizes the signatures of density waves, but does not suppress them completely.
Second, we mention briefly how the dynamical cluster approximation (DCA), another cluster extension of dynamical mean-field theory, circumvents the problem and at what cost. 
Third, we discuss the various periodization schemes suggested to restore translation invariance and full momentum dependence in CDMFT~\cite{senechal_spectral_2000,kotliar_cellular_2001,biroli_cluster_2002,stanescu_fermi_2006,sakai_cluster-size_2012}.
Among these schemes, the Green's function periodization and the cumulant periodization are usually preferred. The results of this paper show that the Green's function periodization retains signatures of the broken translation invariance in a way fully consistent with the signatures seen in the unperiodized Green's function.
As we explain in Sec.~\ref{sec_periodization}, this is because the Green's function periodization corresponds to the diagonal components of the actual CDMFT results in momentum space.
The cumulant periodization, in contrast, results in a new Green's function object, fundamentally different from the unperiodized Green's function, and better seen as an \emph{estimator} of the actual lattice solution, as originally suggested~\cite{kotliar_cellular_2001, biroli_cluster_2002, stanescu_cellular_2006, stanescu_fermi_2006}.
Whether the cumulant periodization actually supresses the effects of broken translation invariance, or merely deforms them, remains an open question.
Note that unphysical results with the cumulant periodization have been reported~\cite{simard_master}\footnote{The superfluid stiffness calculated with a cumulant-periodized Green’s function develops unphysical minima in its doping dependence (see Fig. 6.11 of Ref.~\cite{simard_master}). This does not happen with the Green’s function periodization or when no periodization is used~\cite{simard_superfluid_2019}.}. 

The paper is divided as follows. Section~\ref{sec_hubbard} defines the model. Section~\ref{sec_cdmft} reviews some key aspects of CDMFT. Section~\ref{sec_translation} describes how CDMFT breaks translation invariance and contrasts this with DCA. Section~\ref{sec_periodization} reviews the various periodization schemes and explains why Green's function periodization allows to study the breaking of translation invariance, whereas cumulant periodization does not. Section~\ref{spectral} defines the spectral quantities we study. Section~\ref{sec_order} defines the order parameters for superconductivity and the cluster-shaped density waves. Section~\ref{sec_meanfield} presents the mean-field model we use to reproduce low-energy CDMFT features. The rest of the paper presents the results, with Sec.~\ref{sec_intrinsic} establishing the main signatures of the density waves and Sec.~\ref{sec_doping} presenting an analysis of the doping dependence of these signatures. Finally, Sec.~\ref{experiments} discusses the relation of these results with regards to experiments, before concluding.

\section{Models \& Methods}
\subsection{The Hubbard model}\label{sec_hubbard}
We start from the Hubbard model $\hat{H} = \hat{t}+\hat{U}$ on a square lattice, with the kinetic term
\begin{align}
\hat{t} &=
-\sum_{\vec r,\vec a,\sigma}t_{\vec a}
c^\dag_{\vec r+\vec a,\sigma}
c^{\phantom{\dag}}_{\vec r,\sigma}
-
\mu\sum_{\vec r,\sigma}c^\dag_{\vec r,\sigma}
c^{\phantom{\dag}}_{\vec r,\sigma},
\label{kinetic}
\end{align}
and local interaction
\begin{align}
\hat{U} &=
\sum_{\vec r}Uc^\dag_{\vec r,\uparrow}c^{\phantom{\dag}}_{\vec r,\uparrow}c^\dag_{\vec r,\downarrow}c^{\phantom{\dag}}_{\vec r,\downarrow}.
\end{align}
Operators $c_{\vec r,\sigma}^\dagger$ and $c_{\vec r,\sigma}$ respectively create and annihilate electrons of spin $\sigma$ at positions $\vec r$, and vectors $\vec a$ point to neighbouring sites of $\vec r$. Hole doping $p$ is controlled through the chemical potential $\mu$. We use band parameters $t=1$, $t'=-0.3$ and $t''=0.2$ for first, second, and third-neighbour hopping, respectively~\cite{andersen_lda_1995} and we set $U=8$. Those parameters are suggested by electronic-structure calculations~\cite{pavarini_band-structure_2001} and by the good qualitative agreement found with cuprates~\cite{kancharla_anomalous_2008, foley_coexistence_2019}.

\subsection{CDMFT}\label{sec_cdmft}

The purpose of this section is to define our notation and to review a few key aspects of CDMFT. For a full description of the method, we refer the reader to Refs.~\onlinecite{maier_quantum_2005},~\onlinecite{kancharla_anomalous_2008},~\onlinecite{senechal_introduction_2008}, and~\onlinecite{senechal_quantum_2015}

CDMFT provides an approximate solution to the Hubbard model, in the form of a cluster Green's function $G^{\text{c}}$ and a lattice Green's function $G^{\text{L}}$ that satisfy the following self-consistency condition:
\begin{align}
G^{\text{c}}_{\vec R,\vec R'}(z)
=
\frac{N_c}{N}\sum_{\tilde{\vec k}}
G^{\text{L}}_{\vec R,\vec R'}(\tilde{\vec k},z)
\label{cdmft}.
\end{align}
In our notation, $z$ is the complex frequency ($z=i\omega_n$ for the Matsubara Green's function or $z=\omega+i\eta$ for the retarded Green's function). Positions are expressed as $\vec r = \tilde{\vec r}+\vec R$, with $\tilde{\vec r}$ denoting a superlattice vector (the position of the cluster) and $\vec R$ the position within the cluster. Accordingly, wave vectors in the Brillouin zone are written as $\vec k=\tilde{\vec k} +\vec K$, with $\vec K$ a reciprocal super-lattice wave vector ($\vec K\cdot\tilde{\vec r}=2\pi$) and $\tilde{\vec k}$ restricted to the reduced Brillouin zone. All quantities are thus expressed in an intermediate Fourier representation $(\vec R,\tilde{\vec k})$ 
with $N_c$ the number of cluster sites and $N$ the total number of lattice sites.

A matrix representation in the coordinates $\vec R,\vec R'$ is necessary if we want to write inverse Green's functions. We denote these matrices with bold typeface. In this representation, 
the lattice Green's function $G^{\text{L}}_{\vec R,\vec R'}(\tilde{\vec k},z)$, or $\vec G^{\text{L}}(\tilde{\vec k},z)$, in matrix notation, is given by 
\begin{align}
\vec G^{\text{L}}(\tilde{\vec k},z) =
\left[
z - \vec t(\tilde{\vec k}) - \vec \Sigma^\text{c}(z)
\right]^{-1},
\label{lattice_green}
\end{align}
where $\vec t(\tilde{\vec k})$ is the kinetic term~\eqref{kinetic} in the $(\vec R,\tilde{\vec k})$ matrix representation and $\vec \Sigma^\text{c}(z)$ is the cluster self-energy. The use of the cluster self-energy $\vec \Sigma^\text{c}(z)$ in place of the true lattice self-energy is the main approximation of CDMFT. To find $\vec \Sigma^\text{c}(z)$, an Anderson impurity model, defined by $\hat{H}^{\text{AIM}}=\hat{t}^{\text{c}}+\hat{U}+\hat{\theta}+\hat{\epsilon}$ must be solved, where $\hat{t}^{\text{c}}+\hat{U}$ is the restriction of the Hubbard model to a cluster, and $\hat{\theta}+\hat{\epsilon}$ is a non-interacting environment to the cluster. Namely, $\hat{\theta}$ is a hopping Hamiltonian from cluster to non-interacting medium called the \emph{bath}, and $\hat{\epsilon}$ is the Hamiltonian of this bath. The numerical solution of the impurity model yields the cluster Green's function,
\begin{align}
\vec G^{\text{c}}(z) =
\left[
z - \vec t^{\text{c}} - \vec \Sigma^{\text{c}}(z) - \vec \theta\frac{1}{z-\vec \epsilon}\vec\theta^{\dag} 
\right]^{-1},
\label{cluster_green}
\end{align}
from which the self-energy $\vec \Sigma^\text{c}(z)$ is extracted.
Here, $\vec \theta$ and $\vec \epsilon$ are the matrix representations of $\hat{\theta}$ and $\hat{\epsilon}$, respectively. The latter are adjusted iteratively to reach the self-consistency condition~\eqref{cdmft}. 

The fine details of our CDMFT implementation can be found in Kancharla et al.~\cite{kancharla_anomalous_2008}. In brief, we use an exact diagonalization impurity solver on a $2\times2$ cluster with 8 bath orbitals at zero temperature. The only difference with Ref.~\onlinecite{kancharla_anomalous_2008} is that our distance function uses a sharp cutoff at~$\omega_c=2$, as studied in Ref.~\onlinecite{senechal_bath_2010}, instead of~$\omega_c=1.5$.
Note that we discuss the effects of using a finite bath at Sec.~\ref{sec_unperiodized} and Sec.~\ref{sec_eta}

\subsection{Translation Invariance}\label{sec_translation}
\begin{figure}
\includegraphics[width=3.5cm]{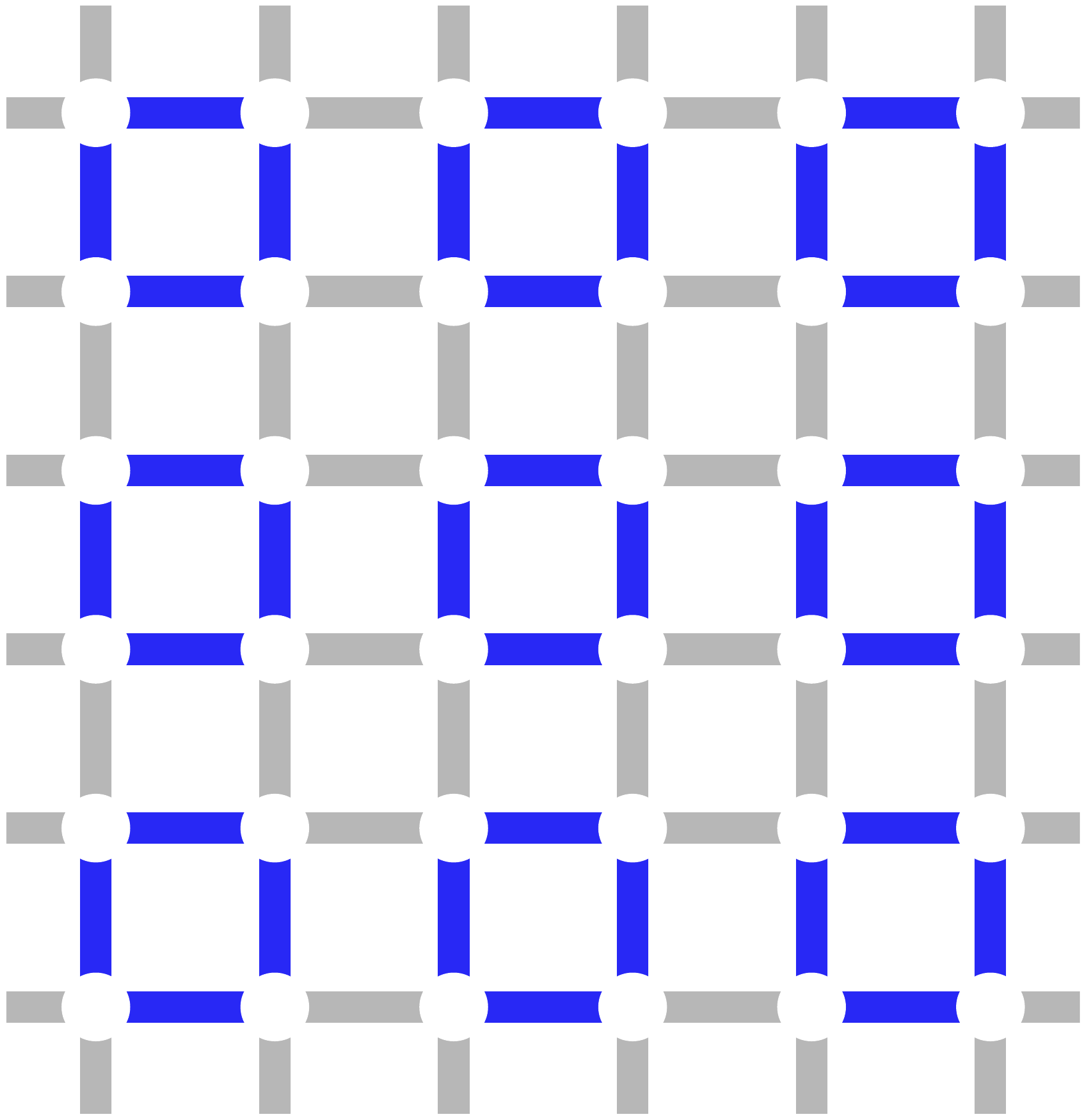}
\caption{$2\times2$ tiling realized by the self-energy of the CDMFT lattice Green's function $\vec G^{\text{L}}(\tilde{\vec k})$. This pattern is also the one of a bond-centered charge density waves with first-neighbour hoppings $t_{\hat{\textbf x}}$ and $t_{\hat{\textbf y}}$ respectively oscillating according to wave vectors $\vec Q_x = \pi \hat{\vec x}$ and $\vec Q_y = \pi \hat{\vec y}$. We call the density waves following this pattern \emph{cluster density waves} (clus.-DW).}
\label{fig_22tilling}
\end{figure}

CDMFT breaks translation invariance. The hopping matrix $\vec{t}(\tilde{\vec k})$, is expressed as~\cite{senechal_quantum_2015}:
\begin{align}
t_{\vec R,\vec R'}(\tilde{\vec k})
=
\sum_{\tilde{\vec r}'}
e^{i\tilde{\vec k}\cdot\tilde{\vec r}'} t_{\vec R,\tilde{\vec r}'+\vec R'},
\label{cdmftRep}
\end{align}
and the corresponding cluster hopping Hamiltonian is obtained as $\vec t^c = \frac{N_c}{N}\sum_{\tilde{\vec k}}\vec t(\tilde{\vec k})$ and has open boundary condition. The self-energy is taken to be zero between clusters, as depicted in Fig.~\ref{fig_22tilling}. As we will discuss, this is the direct cause of the density waves studied in this paper.

The method known as dynamical cluster approximation (DCA)~\cite{hettler_nonlocal_1998,hettler_dynamical_2000}, also a cluster extension of DMFT, is almost identical to CDMFT (see Ref.~\onlinecite{maier_quantum_2005}), but it preserves translation invariance. As first explained in Ref.~\onlinecite{biroli_cluster_2002}, DCA can be entirely expressed in the CDMFT formalism with the hopping matrix $\vec{t}^{\text{DCA}}(\tilde{\vec k})$ expressed as
\begin{align}
t^{\text{DCA}}_{\vec R,\vec R'}(\tilde{\vec k})
=
e^{-i\tilde{\vec k}\cdot(\vec R-\vec R')}
\sum_{\tilde{\vec r}'}
e^{i\tilde{\vec k}\cdot\tilde{\vec r}'}
t_{\vec R,\tilde{\vec r}'+\vec R'},
\label{dca}
\end{align}
leading to a different cluster hopping Hamiltonian $\vec t^{\text{c},\text{DCA}} = \frac{N_c}{N}\sum_{\tilde{\vec k}}\vec t^{\text{DCA}}(\tilde{\vec k})$, which is periodic in $\vec R$ space. As a consequence, momentum $\vec K$ is a good quantum number of the impurity system and DCA preserves translation invariance. For this reason, none of the density waves discussed in this paper will be found in DCA.

Although DCA preserves translation invariance, it comes at the cost of a piecewise constant self-energy in the Brillouin zone: the DCA lattice self-energy is defined only for the few $\vec K$ momenta of the periodic cluster. By contrast, in CDMFT, the phase $e^{-i\tilde{\vec k}\cdot(\vec R-\vec R')}$, which makes $\vec t^{\text{c,DCA}}$ periodic, is not part of the impurity model, and so it can be restituted to the self-energy when returning to $\vec k$-space. This allows to bring back full $\vec k$-dependence, as explained in the next section. We do not consider DCA further in this paper.

\subsection{Periodization}\label{sec_periodization}
To estimate the full $\vec k$-dependence of the CDMFT Green's function~\eqref{lattice_green}, we need to perform periodization. 
This aspect of CDMFT has already been extensively studied~\cite{senechal_spectral_2000, biroli_cluster_2002,stanescu_fermi_2006,sakai_cluster-size_2012}. Here we revisit periodization of an arbitrary quantity $\vec Q(\tilde{\vec k})$ to highlight that it consists of two steps:
\begin{enumerate}
\item Change the basis from $(\vec R, \vec R',\tilde{\vec k})$-space to $(\vec k,\vec k')$-space.
\item Discard the off-diagonal elements in $(\vec k,\vec k')$-space.
\end{enumerate}
The first step is the following unitary transformation:
\begin{align}
Q(\tilde{\vec k}+\vec K,\tilde{\vec k}+\vec K') 
&= 
\frac{1}{N_c}\sum_{\vec R,\vec R'} e^{-i\left((\tilde{\vec k}+\vec K)\cdot\vec R - (\tilde{\vec k}+\vec K')\cdot\vec R'\right)} Q_{\vec R,\vec R'}(\tilde{\vec k}).
\label{unitaryk}
\end{align}
which preserves all information.
The second step amounts to setting $Q(\tilde{\vec k}+\vec K,\tilde{\vec k}+\vec K')=0$ for all $\vec K'\neq\vec K$, which discards information and results in a new object $Q(\vec k = \tilde{\vec k}+\vec K)$
\begin{align}
Q(\vec k) 
&= 
\frac{1}{N_c}\sum_{\vec R,\vec R'} e^{-i\vec k\cdot(\vec R - \vec R')} Q_{\vec R,\vec R'}(\vec k),
\label{periodization}
\end{align}
where $\vec k$ belongs to the original Brillouin zone. The last equation is the definition of periodization as originally proposed in Ref.~\onlinecite{senechal_spectral_2000}; we stress again that it implicitly discards the off-diagonal elements in $(\vec k,\vec k')$. This can be explicitely written as $Q(\vec k) = \delta_{\vec k,\vec k'}Q(\vec k,\vec k')$

The three typical candidates for $\vec Q(\tilde{\vec k})$ are the lattice Green's function $\vec G^{\text{L}}(\tilde{\vec k},z)$, the self-energy $\vec \Sigma^c$, or the cumulant $\vec M(z) \equiv (z + \mu - \vec\Sigma^{\text{c}})^{-1}$. 
To compare the three resulting approximations, we can write their respective $\vec k$-dependent Green's function. They are known as (i) the G-periodization~\cite{senechal_spectral_2000}:
\begin{align}
G(\vec k,z) = \frac{1}{N_c}\sum_{\vec R,\vec R'} e^{-i\vec k(\vec R - \vec R')}
\left[\frac{1}{z - \vec t({\vec k}) - \vec \Sigma^{\text{c}}(z)}\right]_{\vec R,\vec R'},
\label{green_periodization}
\end{align}
(ii) the $\Sigma$-periodization~\cite{kotliar_cellular_2001}:
\begin{align}
G^{\Sigma}(\vec k,z) \equiv
\left[
z - t(\vec k) 
-
\frac{1}{N_c}\sum_{\vec R,\vec R'} e^{-i\vec k(\vec R - \vec R')}
\left[\vec\Sigma^{\text{c}}(z)\right]_{\vec R,\vec R'}
\right ]^{-1},
\label{sigma_periodization}
\end{align}
and (iii) the M-periodization~\cite{stanescu_fermi_2006}:
\begin{align}
G^{\text{M}}(\vec k,z) \equiv
\left[
\left(
\frac{1}{N_c}\sum_{\vec R,\vec R'} e^{-i\vec k(\vec R - \vec R')}
\left[
\frac{1}{z +\mu - \vec\Sigma^{\text{c}}(z)}
\right]_{\vec R,\vec R'}
\right)^{-1}
\hspace{-2mm}- \epsilon(\vec k)
\right ]^{-1}.
\label{cumulant_periodization}
\end{align}
where $t(\vec k)=\epsilon(\vec k)-\mu$ is the non-interacting dispersion, i.e., the $\vec k$-space representation of the operator $\hat{t}$. 

In the rest of this paper, we use only the G-periodization. The reasons for this are explained below.

First, note that periodized quantities \emph{do not} satisfy the DMFT self-consistency condition~\eqref{cdmft}.
Indeed, the quantities that do satisfy the self-consistency condition are $G^{\text{L}}_{\vec R,\vec R'}(\tilde{\vec k},z)$, $M_{\vec R,\vec R'}(z) $, and $\Sigma^{c}_{\vec R,\vec R'}(z)$. With the unitary transform~\eqref{unitaryk}, we can transform these respectively to $G^{\text{L}}(\vec k,\vec k',z)$, $M(\vec k,\vec k',z)$, and $\Sigma(\vec k,\vec k',z)$, which are the same objects represented in $\vec k$-space, and therefore also satisfy the self-consistency condition. However, they are not diagonal in $\vec k$. With the off-diagonal elements removed through periodization~\eqref{periodization}, $G(\vec k,z)$, $M(\vec k,z)$, and $\Sigma(\vec k,z)$ become different objects which are now diagonal in $\vec k$ but which do not satisfy the self-consistency condition\footnote{Periodized-CDMFT (PCDMFT)~\cite{biroli_cluster_2004} includes the periodized self-energy in the self-consistent condition. This, in fact, corresponds to build a new self-consistency condition in order to preserve translation invariance. Such a self-consistency condition is not as clearly related to the one of DMFT as those of DCA and CDMFT, because the self-energy used in the lattice is not an impurity self-energy. In the 1D case, PCDMFT performs worst than plain CDMFT in the Mott insulating regime~\cite{capone_cluster-dynamical_2004}.}.

Second, even though $G(\vec k,z)$, $M(\vec k,z)$ and $\Sigma(\vec k,z)$ are all translation invariant by construction, they are no more than the diagonal elements of $G^{\text{L}}(\vec k,\vec k',z)$, $M(\vec k,\vec k',z)$, and $\Sigma(\vec k,\vec k',z)$, which all break translation invariance. As a consequence, these diagonal elements (the periodized quantities) preserve signatures of broken translation invariance (see appendix~\ref{a1}). Section~\ref{section_periodized} studies these signatures of broken translation invariance in the G-periodized spectral weight.

These two arguments alone---that periodized quantities do not satisfy the DMFT self-consistency condition, and that they preserve signatures of broken translation invariance---do not justify our preference for the G-periodization: they apply equally to all periodization schemes.

What makes the G-periodized Green's function preferable is its relation to the truly self-consistent Green's function $G^{\text{L}}(\tilde{\vec k}+\vec K,\tilde{\vec k}+\vec K',z)$. 
Indeed, the $\Sigma$- and M-periodized Green's function, given by Eqs.~\eqref{sigma_periodization} and~\eqref{cumulant_periodization}, discard off-diagonal elements before taking matrix inversions. This can be seen from the sums on $\vec R,\vec R'$, coming from Eq.~\eqref{periodization}, and which are taken within inverse operations\footnote{If the off-diagonal elements were kept in matrix inversions, all schemes would be equivalent, and would lead to $G^{\text{L}}(\tilde{\vec k}+\vec K,\tilde{\vec k}+\vec K',z)$, simply being different implementations of the same unitary transformation~\eqref{unitaryk}.}. Therefore, the resulting $G^{\Sigma}(\vec k,z)$ and $G^{\text{M}}(\vec k,z)$ have no evident relation to the self-consistent Green's function $G^{\text{L}}(\tilde{\vec k}+\vec K,\tilde{\vec k}+\vec K',z)$. By contrast, in the G-periodized $G(\vec k,z)$, the off-diagonal elements are discarded outside all inversions, meaning that $G(\vec k,z)$ simply corresponds to the diagonal elements of $G^{\text{L}}$, i.e., $G(\vec k,z) = \delta_{\vec K,\vec K'}G^{\text{L}}(\tilde{\vec k}+\vec K,\tilde{\vec k}+\vec K',z)$.

Therefore, $G(\vec k,z)$ can be seen as an economical way to study the diagonal elements of $G^{\text{L}}(\vec k,\vec k',z)$. Equivalently, if we wanted to study the diagonal elements of $M(\vec k,\vec k',z)$ and $\Sigma(\vec k,\vec k',z)$, we could consider the periodized $M(\vec k,z)$ and $\Sigma(\vec k,z)$. However, using the latter to construct a new Green's functions transforms the information about translation invariance in a very uncontrolled manner (referring to off-diagonal elements being discarded before matrix inversions). The resulting Green's function, $G^{\Sigma}(\vec k,z)$ and $G^{\text{M}}(\vec k,z)$, should therefore be considered as estimators (as originally suggested in Refs.~\onlinecite{kotliar_cellular_2001,biroli_cluster_2002, stanescu_cellular_2006, stanescu_fermi_2006}) of the hypothetical translation invariant lattice Green's function. By contrast, the G-periodized $G(\vec k,z)$ is not an estimator; it is the diagonal part, in $\vec k$-space, of the actual CDMFT solution.

The G-periodized Green's function can thus be useful to study the actual CDMFT solution. As we explain in the next section, translation invariant one-particle observables can be computed using only a trace involving the Green's function, \emph{i.e.}, its diagonal elements (this is clear when one uses the Luttinger-Ward functional to derive the CDMFT formalism~\cite{chitra2001effective,potthoff2003self,tremblay_pseudogap_2006}). For such one-particle observables, the G-periodized Green's function can be used to identify the momentum space structures, contained in the CDMFT solution, which give rise to these local quantities. Section \ref{section_periodized} provides a good example of this: we explain the origins of particular features (subgap structures) found in the local density of states by studying the G-periodized spectral weight.




\subsection{Spectral quantities}\label{spectral}

The local density of states is computed from the unperiodized Green's function using 
\begin{align} 
N(\omega) 
&= 
\frac{N_c}{N}\sum_{\tilde{\vec k}} 
\text{tr}\left[
-\frac{1}{\pi}\text{Im}\left\{\vec G^{\text{L}} (\tilde{\vec k},\omega+i\eta)\right\}
\right],
\end{align}
where the trace is over cluster indices. The Lorentzian broadening $\eta$ is specified for each case discussed below. Note that using the G-periodized Green's function would yield the same result, since the trace is invariant under unitary transformations and only requires the diagonal elements. The G-periodized spectral weight is given by:
\begin{align}
A(\vec k,\omega)
&=
-\frac{1}{\pi}\text{Im}\left\{
G (\vec k,\omega+i\eta)
\right\},
\label{g_spectral}
\end{align}
with $G(\vec k,z)$ given by Eq.~\eqref{green_periodization}. 

When studying superconductivity, we will also consider the zero-temperature $\vec k$-resolved gap function:
\begin{align}
F(\vec k) =
 \int\limits^{\infty}_{-\infty}\frac{\text{d}\omega}{2\pi i} F(\vec k,i\omega),
\label{gorkov}
\end{align}
where $F(\vec k,z)$ is the Gor'kov function, also known as the anomalous part of $G(\vec k,z)$. Integrating this function over Matsubara frequency at zero temperature, as in~\eqref{gorkov}, yields the pairing amplitude at each wave vector.

\subsection{Order Parameters}\label{sec_order}
We consider two types of CDMFT solutions to the Hubbard model: a normal solution and a superconducting solution.
In the normal solution, no spontaneous symmetry breaking is allowed, whereas in the superconducting solution, superconductivity is allowed and it develops for certain values of interaction $U$ and doping $p$. It is the first, the normal solution, which is associated with the pseudogap state above $T_c$, although here it is obtained at zero temperature by preventing broken symmetry.

To probe the orders present in the CDMFT solutions, we compute the associated order parameters.
For example, the pairing operator of $d$-wave superconductivity (SC) is
\begin{align}
\hat{\Delta} = 
&\sum_{\vec r,\vec a}
\Delta_{\vec a}
\left(
c_{\vec r+\vec a,\uparrow}c_{\vec r,\downarrow}-c_{\vec r+\vec a,\downarrow}c_{\vec r,\uparrow}
\right)+\text{H.c.}
\label{sc_op}
\end{align}
with $\Delta_{\hat{\textbf x}}=-\Delta_{\hat{\textbf y}}=1$. This yields the usual \emph{d}-wave form factor $\Delta(\vec{k})=(\cos k_x - \cos k_y)$ in $\vec k$-space.
In this paper, we are particularly interested in the density-wave operator
\begin{align}
\hat{t}_{Q}
=&\sum_{\vec r,\vec a,\sigma}\sum_{\vec Q}
 t_{\vec Q,\vec a}
 {e}^{i\vec Q\vec r}
 c^\dagger_{\vec r+\vec a,\sigma}c^{\phantom{\dag}}_{\vec r,\sigma}
 +\text{H.c.},
\label{bdw_op}
\end{align}
where $t_{\vec Q_x,\hat{\textbf x}}=t_{\vec Q_y,\hat{\textbf y}}=1$ for two wave vectors $\vec Q_x = \pi \hat{\vec x}$ and $\vec Q_y = \pi \hat{\vec y}$. These parameters yield a density wave that reproduces the 2$\times$2 cluster tiling used in our CDMFT scheme, as illustrated in Fig.~\ref{fig_22tilling}. Note that it is a bond-centered density wave. 
We are also interested in the pair-density-wave operator
\begin{align}
\hat{\Delta}_{Q}
=\sum_{\vec r,\vec a,\vec Q}
 &\Delta_{\vec Q,\vec a}
 {e}^{i\vec Q\vec r}
\left(
c_{\vec r+\vec a,\uparrow}c_{\vec r,\downarrow}-c_{\vec r+\vec a,\downarrow}c_{\vec r,\uparrow}
\right)+\text{H.c.}
\label{pdw_op}
\end{align}
where $\Delta_{\vec Q_x,\hat{\textbf x}}=-\Delta_{\vec Q_y,\hat{\textbf y}}=1$ that yields the same 2$\times$2 cluster-shaped tiling for Cooper pairs. This pattern for charge and pair density waves is what we mean by ``cluster density waves'' (clus.-DW).
Note that bond-centered density waves similar to~\eqref{bdw_op} were considered as models of the charge order observed in cuprates~\cite{allais_connecting_2014,verret_subgap_2017}. However, the periodicity considered was close to 4 unit cells, with quasi-incommensurate $\vec Q$, as observed in cuprates, instead of the commensurate periodicity of 2 unit cells considered here.

To compute the expectation values of these operators~\cite{senechal_quantum_2015}, we use the unperiodized lattice Green's function
\begin{equation}
\langle \hat{O}\rangle = \frac{N_c}{N}\sum_{\tilde{\vec k}}\int\limits_{-\infty}^{\infty}\frac{\text{d}\omega}{2\pi} \text{tr}\left[ \vec{O}(\tilde{\vec k})\vec{G}^{\text{L}}(\tilde{\vec k},\omega)\right].
\label{ave_lat}
\end{equation}
We use $\vec G^{\text{L}}(\tilde{\vec k},z)$ because it is the only one that can confirm the presence of the cluster density waves. Indeed, the cluster Green's function, $\vec G^{\text{c}}(z)$, lacks the relevant inter-cluster elements, whereas any periodized Green's function is deprived from the relevant off-diagonal elements $\langle c^{\phantom\dag}_{\vec k+\vec Q} c^\dag_{\vec k}\rangle$.

\subsection{Mean field to simulate cluster density waves}\label{sec_meanfield}
Finally, as demonstrated in Sect.~\ref{sec:results} below, we can reproduce the superconducting CDMFT spectrum at low energy with the following phenomenological mean-field model for $d$-wave superconductivity coexisting with cluster density waves:
\begin{equation}
\hat{H}^{\text{MF}} = \hat{t} + D\hat{\Delta} +B\hat{t}_{Q} + P\hat{\Delta}_{Q}.
\label{mf}
\end{equation}
With $B=0$ and $P=0$ this models become a $d$-wave BCS model with amplitude of the superconducting gap controlled by $D$. When present, the amplitude of the bond-centered charge density waves is controlled by $B$ and that of the pair density wave by $P$.

Such a mean-field model cannot be used to fit the normal CDMFT solution, because a mean field for the pseudogap present in CDMFT would be necessary. The possible existence of a mean field for the pseudogap is still an active field of research~\cite{yang_phenomenological_2006, lee_amperean_2014, scheurer_topological_2018}, outside the scope of the present paper. By contrast, the superconducting CDMFT solution comes with well-defined low-energy Bogoliubov quasiparticles that are well suited to mean-field modeling. We can therefore study the interplay of these quasiparticles with the cluster density waves. It is curious that a pseudogap mean field is not necessary to reproduce the superconducting results. It might be linked to the cancellation of the pseudogap poles by the superconductivity poles documented in Ref.~\onlinecite{sakai_hidden_2016}.


\section{Results}\label{sec:results}

\subsection{Intrinsic density waves}\label{sec_intrinsic}

\subsubsection{unperiodized results}\label{sec_unperiodized}
\begin{figure}
\vspace{5mm}
\includegraphics[width=\columnwidth]{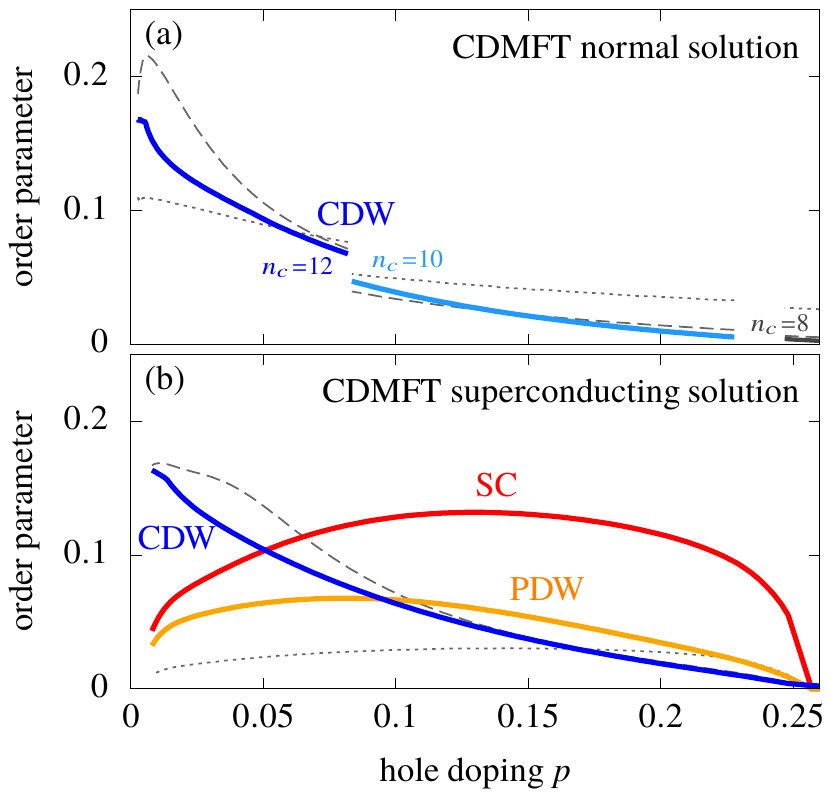}
\caption{
(a) Order parameter $\langle \hat{t}_{Q} \rangle$ of the bond-centered charge density waves (CDW) as a function of doping~$p$ in the normal solution (no spontaneously broken symmetry). This order parameter nearly scales with the zero-frequency nearest-neighbour (nn) cluster self-energy $\tfrac{1}{8}|\Sigma^{c}_{\text{nn}}(z=0)|$ (dashed line) and with its high-frequency second moment $\tfrac{1}{256}|\omega^{2}\Sigma^{c}_{\text{nn}}(i\omega)|_{\omega\rightarrow\infty}$ (dotted line) where $\Sigma^{c}_{\text{nn}}(z) = \Sigma^{c}_{\vec R=\vec 0,\vec R'=\hat{\vec x}}(z)$. 
(b) Order parameters for superconductivity $\langle \hat{\Delta} \rangle$ (SC), bond-centered charge density waves $\langle \hat{t}_{Q} \rangle$ and pair density waves $\langle \hat{\Delta}_{Q} \rangle$ (PDW) as a function of doping~$p$ in the superconducting solution. Here it is compared to the anomalous nearest neighbour self-energy using $\tfrac{1}{8}|\Sigma^{c,\text{ano}}_{\vec 0,\hat{\vec x}}(z=0)|$ (dashed line) and $\tfrac{1}{256}|\omega^{2}\Sigma^{c,\text{ano}}_{1,2}(i\omega)|_{\omega\rightarrow\infty}$ (dotted line) scaled to fit in the plot.}
\label{order_vs_doping}
\end{figure}

Fig.~\ref{order_vs_doping}(a) shows the order parameter $\langle \hat{t}_{Q} \rangle$ of the cluster-shaped charge density waves as a function of doping in the normal CDMFT solution. The value of $\langle \hat{t}_{Q} \rangle$ is non-zero for the whole doping range, a manifestation of broken translation invariance in CDMFT. It is important to remember that, unlike superconductivity, we cannot suppress these density waves by forcing a symmetric solution of the dynamical mean field (the bath). They are unavoidable products of the method.

The magnitude of $\langle \hat{t}_{Q} \rangle$ reflects the magnitude of the self-energy.
In the lattice Green's function $\vec G^{\text{L}}(\tilde{\vec k},z)=[z - \vec t(\tilde{\vec k}) - \vec \Sigma^{\text{c}}]^{-1}$, the real part of the self-energy $\vec \Sigma^{\text{c}}$ acts as a hopping modifier from $t_{\vec r,\vec r'}$ to $t_{\vec r,\vec r'}+\text{Re} \Sigma^{\text{c}}_{\vec r,\vec r'}$. In CDMFT, the first-neighbour self-energy oscillates in space, being 0 between clusters and non-zero within the cluster. This oscillation is what causes the density wave measured by $\langle \hat{t}_{Q} \rangle$.
In Fig.~\ref{order_vs_doping}(a), the value of $\langle \hat{t}_{Q} \rangle$ is largest at half-filling $p=0$, and decreases towards higher dopings. To show that this trend follows the amplitude of the self-energy, we plot the first-neighbour component of the cluster self-energy at zero frequency $\frac{1}{8}|\Sigma^{c}_{1,2}(z=0)|$ and its second moment $\frac{1}{256}|\omega^{2}\Sigma^{c}_{1,2}(i\omega)|_{\omega\rightarrow\infty}$.

We note that $\langle \hat{t}_{Q} \rangle$ undergoes a jump in Fig.~\ref{order_vs_doping}(a), similar to a first order transition.
Such a transition exists in results that use Monte-Carlo solvers with an infinite bath and it constitutes the basis of an explanation for the pseudogap~\cite{sordi_finite_2010, sordi_mott_2011,fratino_organizing_2016}. Evidence of a similar transition was also recently reported using exact diagonalization solvers~\cite{faye_pseudogap--metal_2017, dash2019pseudogap}. However, in our case, the transition is likely triggered by the finite bath of our exact diagonalization solver.
Indeed, two distinct normal solutions are obtained with our implementation of CDMFT: the first lives in the twelve-particle sector of the exactly diagonalized twelve-site cluster-bath model ($n_c=12$), the other lives in the ten-particle sector ($n_c=10$), and the transition we see is between these two sectors.
This probably influences the physics at play, and thus we prefer to ignore this transition in the rest of the paper.
Instead, we will mainly study the superconducting solution, in which the cluster-bath system does not conserve particle number which makes the transition between $n_c=12$ and $n_c=10$ disappear; $\langle \hat{t}_{Q} \rangle$ decreases without discontinuity.

Density waves also appear in the superconducting case. Fig.~\ref{order_vs_doping}(b) shows the order parameters $\langle \hat{t}_Q\rangle$ for the cluster-shaped charge density waves, $\langle \hat{\Delta} \rangle$ for superconductivity, and~$\langle \hat{\Delta}_Q \rangle$ for the cluster-shaped pair density waves as a function of doping in the superconducting CDMFT solution.
First, the value of $\langle \hat{t}_{Q} \rangle$ is non-zero everywhere, largest at half-filling and smoothly decreasing with doping.
Second, the superconducting order parameter $\langle \hat{\Delta} \rangle$ displays the characteristic dome of unconventional superconductivity, as previously reported~\cite{kancharla_anomalous_2008,capone_competition_2006}.
The pair-density-wave order parameter $\langle \hat{\Delta}_Q \rangle$ follows both trends: it traces a dome that decreases faster at high doping. In comparison, the anomalous first-neighbour self-energy is also shown, decreasing with doping, while its second moment traces a dome.

The weakening of the density-wave order parameters with doping highlights that they are a consequences of the (dynamical) mean-field treatment of inter-cluster correlations. Indeed, spatial correlations decrease with doping, and CDMFT becomes equivalent to single-site DMFT, which is translation-invariant.

\begin{figure}
\vspace{5mm}
\includegraphics[width=\columnwidth]{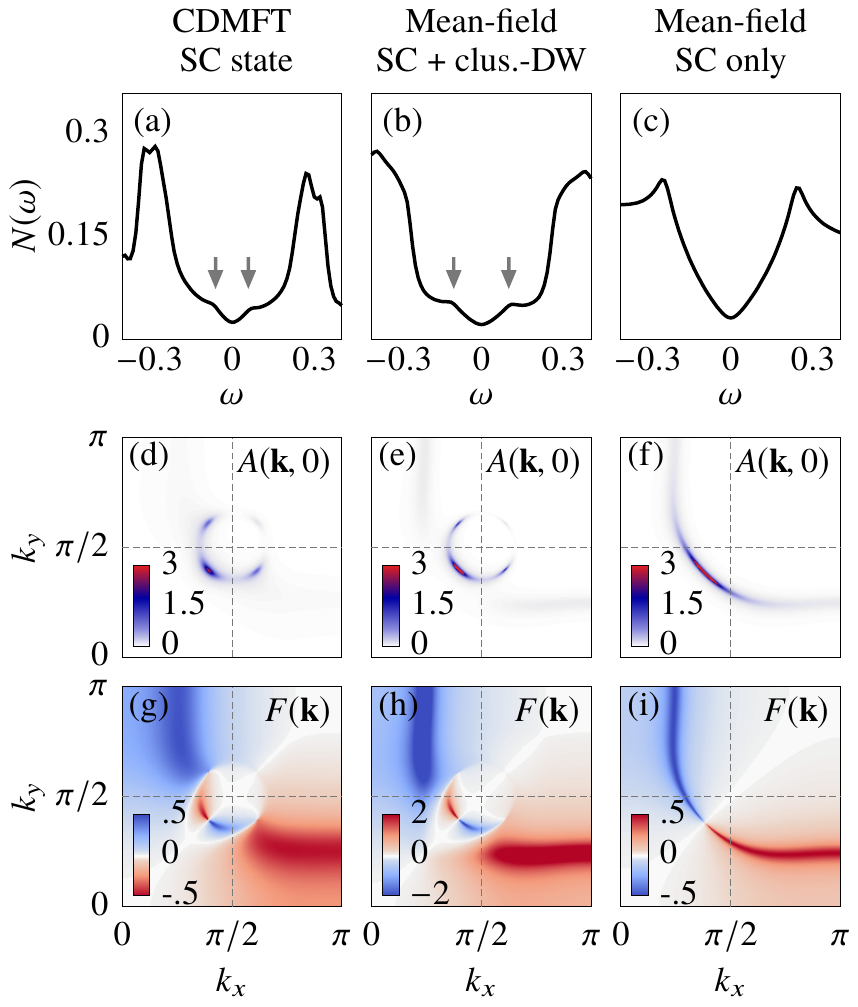}
\caption{ For a broadening $\eta =0.02$:
(a) Density of states $N(\omega)$ in the superconducting CDMFT solution at doping $p=0.08$. The superconducting gap presents subgap structures shown by grey arrows.
(b) Density of states for coexisting superconductivity and density waves in the mean-field model~\eqref{mf} with mean-field values $D=0.07$ for superconductivity, $B=0.2$ for charge density waves, and $P=0.125$ for pair density waves, at chemical potential $\mu=-0.742$ ($p=0.08$). Subgap structures are similar to those in CDMFT.
(c) Density of states for $d$-wave mean-field superconductivity alone, with $B=P=0$, $D=0.07$ and without subgap structures.
(d) Fermi surface (G-periodized spectral weight at $\omega=0$) for the same superconducting CDMFT solution, showing extra nodes at symmetrical position with respect to the Bragg planes of the superlattice (dotted lines).
(e) Fermi surface for the mean-field model with superconductivity and density waves~\eqref{mf}. Extra nodes are similar to those in CDMFT.
(f) Fermi surface for the mean-field model with $d$-wave superconductivity alone, without extra nodes.
(g) $\vec k$-resolved gap \eqref{gorkov} for the superconducting CDMFT solution, displaying three sign changes at symmetrical copies of the nodes.
(h) $\vec k$-resolved gap for the mean-field model with superconductivity and density waves, showing a triple sign change similar to those in CDMFT.
(i) $\vec k$-resolved gap for the mean-field model with $d$-wave superconductivity alone, with only one sign change.
} 
\label{bcs_cdmft}
\end{figure}

We now turn to the consequences of these density waves in spectral quantities. Figure~\ref{bcs_cdmft}(a) shows the density of states in the superconducting CDMFT solution at doping $p=0.08$. The superconducting gap is apparent. Our main observation is that the gap is accompanied by subgap structures: two humps within the gap, on each side of the Fermi level.

In Ref.~\onlinecite{verret_subgap_2017}, similar subgap structures were obtained with mean-field models of $d$-wave superconductivity coexisting with density waves.
Equation~\eqref{mf} is an example of such a model, which we use to reproduce CDMFT results.
Fig.~\ref{bcs_cdmft}(b) and (c) respectively show the superconducting gap of the mean-field model with and without the density waves (setting $B=P=0$ to remove them). This demonstrates that density waves cause subgap structures similar to those found in CDMFT. 
In order to reproduce the superconducting CDMFT results, we adjusted the mean-field values at $D=0.07$ for superconductivity, $B=0.2$ for charge density waves, and $P=0.125$ for pair density waves. These arbitrary mean-field values are acceptable for a comparison with CDMFT because the energy scales (effective $t$ and effective $\omega$) obtained in CDMFT are strongly renormalized by the interaction. If we tried instead to adjust $D$ to the mean-field average $U\langle c_{\vec r,\uparrow} c_{\vec r+\vec a,\downarrow}\rangle$ or to equate the self-energies of both models, this renormalization would cause problems. After all, since density waves \emph{can} reproduce CDMFT results it already suggests that the CDMFT subgap structures are probably caused by the artificial cluster-shaped density waves measured by $\langle \hat{t}_Q\rangle$ and $\langle \hat{\Delta}_Q\rangle$. To confirm this, we next study the Fermi surface.

\subsubsection{G-Periodized results}\label{section_periodized}

The order parameters of Fig.~\ref{order_vs_doping} and the density of states of Fig.~\ref{bcs_cdmft}(a) are obtained from the unperiodized lattice Green's function~\eqref{lattice_green}. By contrast, plotting the Fermi surface requires the $\vec k$ representation of the Green's function. Here we only consider the G-periodized spectral weight \eqref{g_spectral}, for reasons explained in Secs.~\ref{sec_periodization} and~\ref{spectral}. Note that we use a relatively small broadening $\eta =0.02$ in order to clearly reveal the effects of broken translation invariance. Larger values $\eta\sim 0.1$ are typically used and are discussed in the next section.

Fig.~\ref{bcs_cdmft}(d) shows the Fermi surface $A(\vec k,\omega=0)$ in the first quadrant of the Brillouin zone for the superconducting CDMFT solution at doping $p\sim0.08$. Instead of a standard single $d$-wave node, the CDMFT Fermi surface has four nodes. These are symmetrically arranged with respect to the $k_x=\pi/2$ and $k_y=\pi/2$ axes, which are the edges of the reduced Brillouin zone of our superlattice of $2\times2$ clusters. This multiplicity of nodes is not seen in experiments; it is a direct consequence of artificially broken translation invariance in CDMFT. To confirm this, Fig.~\ref{bcs_cdmft}(e) and (f) show the corresponding results for the mean-field model, respectively with and without density waves, revealing that $2\times2$ density waves cause extra-nodes similar to those found in CDMFT. We will soon explain how these nodes are linked to the aforementionned subgap structures in the density of states.

Before we do so, note that the correspondence between the mean-field model and the superconducting CDMFT solution is not restricted to the Fermi surface. Fig.~\ref{bcs_cdmft}(g) shows the energy-integrated Gor'kov function~\eqref{gorkov}, effectively giving the sign and amplitude of the gap as a function of $\vec k$, in the first quadrant of the Brillouin zone. The gap changes sign not once, as expected in a standard $d$-wave picture, but three times, and it has a fairly complex structure. Fig.~\ref{bcs_cdmft}(h) reproduces this structure almost perfectly with the density-wave mean-field model. Note that the necessary ingredient for the multiple sign changes is the pair density waves.\footnote{The mere presence of pair density waves is not sufficient to get the multiple sign changes: Both the superconducting and the pair-density-wave mean-field values had to be strong enough. Moreover, more than three sign changes were also obtained, for other dopings in the case of CDMFT, or with other mean-field values in the mean-field model.}
Fig.~\ref{bcs_cdmft}(i) shows the mean-field results without the density waves. There is only one node, as expected for $d$-wave superconductivity alone.

The density of states, Fermi surface, and $\vec k$-resolved gap are almost identical in the superconducting CDMFT solution and the mean-field model of $d$-wave superconductivity coexisting with $2\times2$ density waves. This leaves little doubt that density-wave physics is present at low energy in the CDMFT superconducting solution.

\begin{figure}
\vspace{5mm}
\includegraphics[width=\columnwidth]{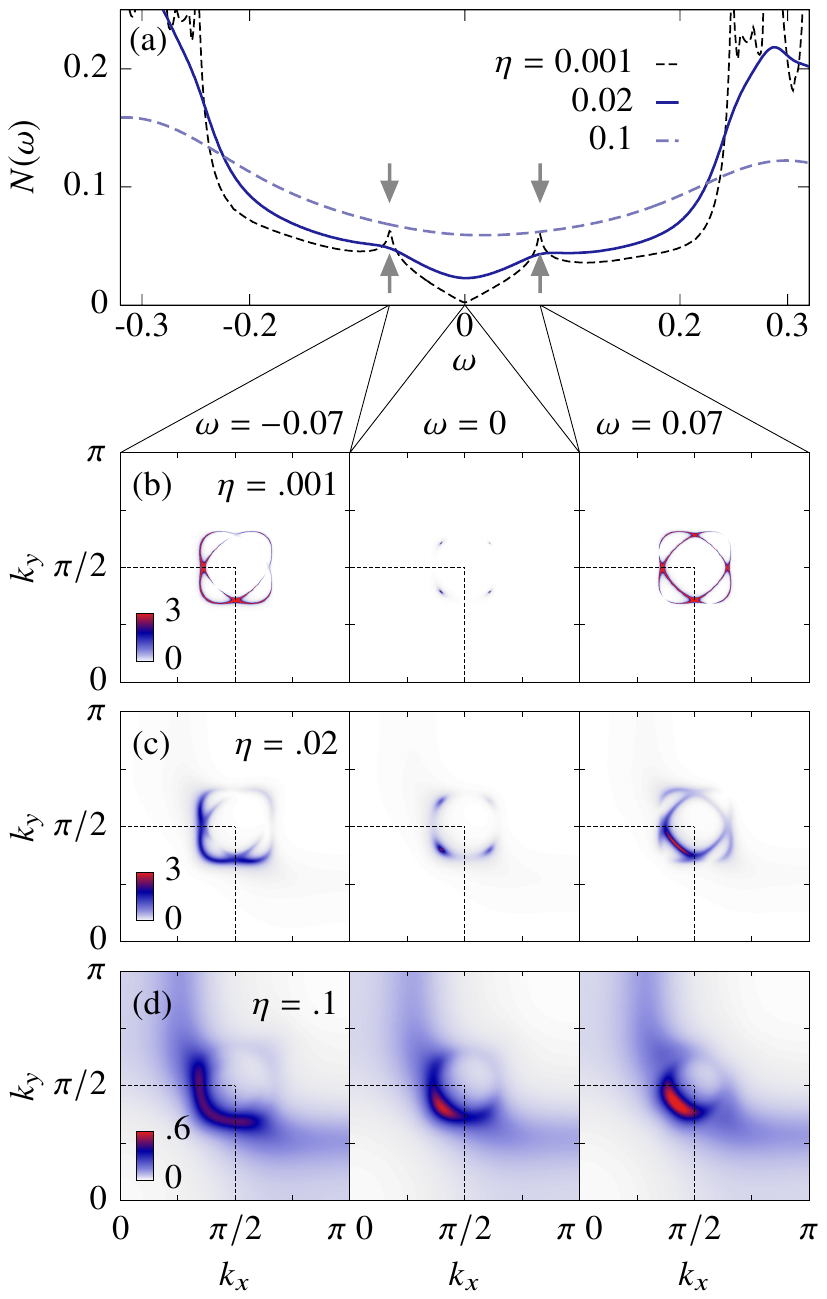}
\caption{
(a) Density of state in the superconducting CDMFT solution (same as in Fig.~\ref{bcs_cdmft}(a) with a close-up of the subgap structure), for three values of Lorentzian broadening $\eta=0.001$, 0.02, and 0.1. 
(b) G-periodized spectral weight at the energy of the subgap structures $\omega=\pm0.07$ (left and right), and at the Fermi energy $\omega=0$ (center), for $\eta=0.001$,
(c) $\eta=0.02$, and
(d) $\eta=0.1$. The dotted lines denote the reduced Brillouin zone boundary.
Note that the central panel is the same as Fig.~\ref{bcs_cdmft}(d).
We know that the superconducting gap goes from $\omega=-0.25$ to $\omega=0.25$ from comparison with the density of state in the normal CDMFT solution and a study of the anomalous part of the Green's function as a function of frequency.}
\label{sc_mdc}
\end{figure}

Now, to understand how broken translation invariance generates the subgap structures, Fig.~\ref{sc_mdc} shows the G-periodized spectral weight at the energy of the subgap structures. A very small Lorentzian broadening $\eta=0.001$ reveals the fine details in the spectral weight of Fig.~\ref{sc_mdc}(b): the subgap structures lie at energies where saddle points (like for van Hove singularities) are present in the quasiparticle spectrum. These happen at the Bragg planes of the superlattice. In other words, the dispersion cone stemming from a given node connects with those of the neighbouring extra nodes at the energy of the subgap structures. This clarifies that the subgap structures and extra nodes are linked, and they are the product of broken translation invariance.

\begin{figure*}
\vspace{5mm}
\includegraphics[width=\textwidth]{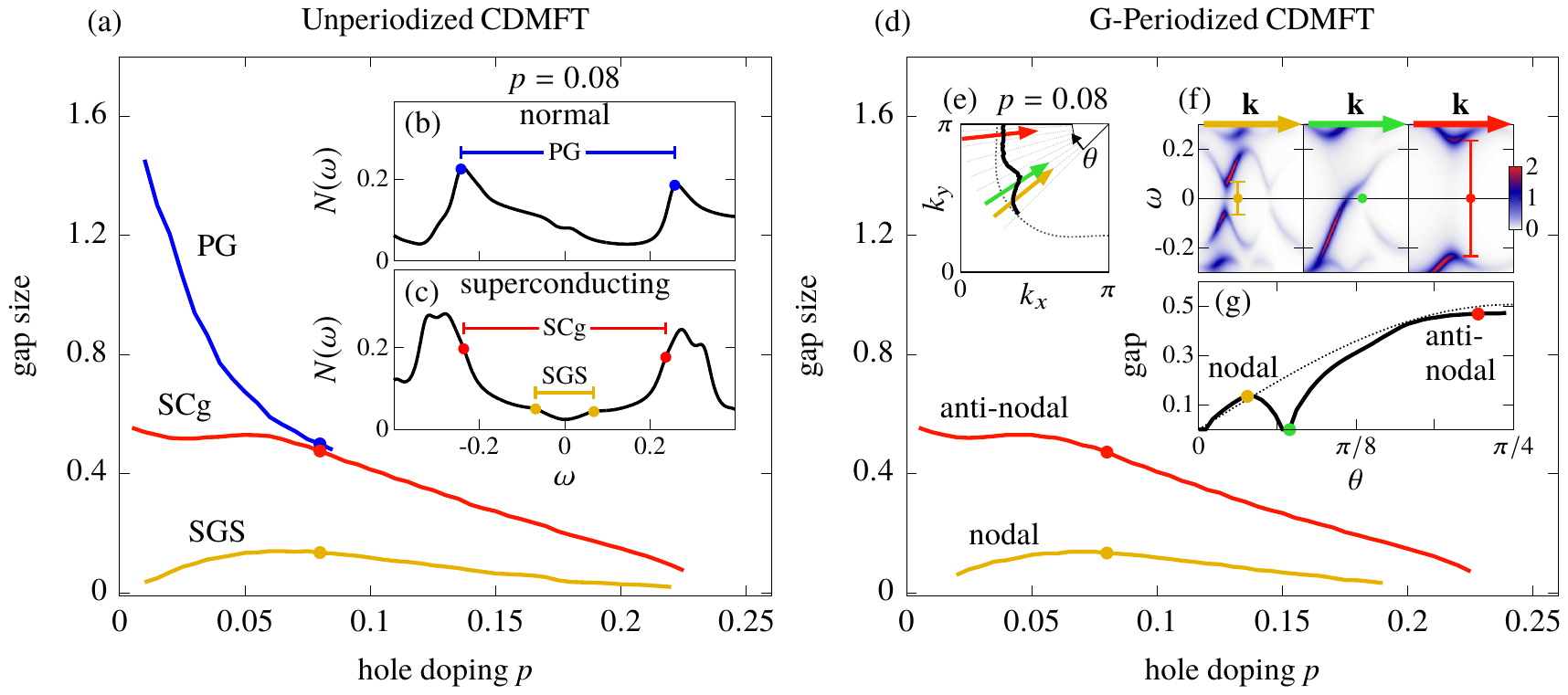}
\caption{
Gaps as a function of doping. (a) Pseudogap (PG), superconducting gap (SCg), and subgap structure (SGS) found in the density of states of the normal and superconducting CDMFT solutions for the unperiodized lattice Green's functions.
(b) An example of how the gaps are identified for doping $p=0.08$ in the density of states $N(\omega)$ of the normal solution and
(c) in the superconducting solution. The gaps are chosen as the distance between two minima in the second derivative of $N(\omega)$.
(d) Nodal and an anti-nodal gap identified in the $\vec k$-dependant spectral weight obtained through G-periodization, and following the same doping dependence as the SCg and SGS.
(e) The black curve gives the position of the gap in the Brillouin zone, identified as the smallest gap along a given direction $\theta$ defined from $(\pi,\pi)$. The angle $\theta=0$ corresponds to the nodal direction.
(f) Along three directions $\theta$ in $\vec k$-space (red, green and golden arrows), results for $A(\vec k,\omega)$ and identification of the smallest gap.
(g) Smallest gaps as a function of $\theta$. The local maximum near $\theta=0$ is identified as the nodal gap, the maximal value near $\theta=\frac{\pi}{4}$ as the anti-nodal gap. The gap returns to zero in between, which is consistent with the extra nodes seen in Fig.~\ref{bcs_cdmft}(d) and~\ref{sc_mdc}(b-c) 
}
\label{gap_vs_doping}
\end{figure*}

\subsubsection{Lorentzian Broadening $\eta$ \& finite bath}\label{sec_eta}

So far we have used a relatively small broadening $\eta =0.02$ compared to the larger values $\eta\sim 0.1$ typically used with these methods~\cite{dahnken_spectral_2002,senechal_cluster_2002,senechal_competition_2005,kancharla_anomalous_2008}.

Using larger values of $\eta$ attenuates certain side effects of using a finite bath. 
The finite bath required by an exact diagonalization solver leads to a large but finite number of eigenvalues for the impurity hamiltonian. 
As a consequence, the cluster Green's function $\vec G^{\text{c}}(z)$ and the self-energy $\vec \Sigma^{\text{c}}(z)$ have discrete poles located close the real axis $\omega$, at a distance $\eta$ (for $z=\omega+i\eta$).
In the density of states $N(\omega)$, this discreteness is washed out when we restore the full lattice dispersion $\vec t(\tilde{\vec k})$ in the lattice Green's function $\vec G^{\text{L}}(\tilde{\vec k},z) = [z-\vec t(\tilde{\vec k})-\vec \Sigma^{\text{c}}(z)]$.
However, the self-energy is still described by a discrete set of poles, and thus
large values of $\eta$ allow to smooth out the resulting discreteness on the real axis $\omega$.

Note that the density waves we study in this paper are caused by the self-energy being zero between clusters, which is unavoidable in CDMFT. Translation invariance is broken irrespectively of the solver used, the self-energy $\omega$-dependence, the size of the bath, or the value of~$\eta$. Therefore, neither a large $\eta$ nor a large bath are expected to remove the density waves studied here.

In the previous sections, a small Lorentzian broadening was used to identify the effects of the cluster density waves. This is because although large $\eta$ cannot remove the density waves, they can hide their consequences to some extent. For example, Fig.~\ref{sc_mdc}(a) shows the density of states for $\eta=0.001$, $0.02$, and $0.1$, revealing that subgap structures disappear for $\eta=0.1$. Fig.~\ref{sc_mdc}(b)~to~(d) show the corresponding spectral weight at the Fermi energy, and at the energies of the subgap structures, revealing that for $\eta=0.1$, the multiple nodes all blend together.
The relative positions of the poles of the Green's function are independent from the value of $\eta$, and thus, although a large value of $\eta$ indeed blurs the spectral weight and hides artifacts, it does not suppress them completely. This can be seen in the middle panel of Fig.~\ref{sc_mdc}(d), at $\eta=0.1$ and $\omega=0$ (similar to the results of the variational cluster approximation in Ref.~\cite{senechal_competition_2005}). The nodes merge to form a pocket around $(\pi/2,\pi/2)$ with very low intensity on one side. Such a pocket is not expected for a d-wave superconducting solution with high energy resolution at zero temperature. A single $d$-wave node should appear instead. With large $\eta$, by focusing on the gross features of the spectrum, one can surmise that without the artificial density waves, there would indeed be a node where it is expected.

\subsection{Doping and Angle Dependence of the Gaps}\label{sec_doping}

The analysis of Figs~\ref{bcs_cdmft} and~\ref{sc_mdc} were only done for doping $p=0.08$. To demonstrate that our conclusions hold at all dopings, and to deepen our understanding of the cluster density waves, this section studies the gaps and the $\vec k$ dependence of the gaps, as a function of doping.

Fig.~\ref{gap_vs_doping}(a) shows the doping dependence of the main gaps found in our CDMFT solutions. The gaps are measured directly in the density of states (unperiodized result) from the distance between two humps in Fig.~\ref{gap_vs_doping}(b) and Fig.~\ref{gap_vs_doping}(c) for $p=0.08$ (humps are found as minima in the second derivatives). The gap identified as the pseudogap (PG) is found in the $n_c=12$ normal CDMFT solution. Fig.~\ref{gap_vs_doping}(a) shows that this gap starts very large near half-filling, decreases rapidly with doping and is absent from the $n_c=10$ solution found at higher dopings. In comparison, the superconducting gap (SCg), found in the superconducting CDMFT solution, stays constant in the underdoped region
and then decreases in the overdoped region. Finally, the subgap structure (SGS) is also measured. It follows a dome-like shape as a function of doping.\footnote{Although this dome-like doping dependence for the SGS might suggest that the SGS is the true superconducting gap, our attempts to separate the small gap from the large gap failed. Both gaps come from the anomalous part of the Green's function. The small subgap cannot be obtained alone.}

Figure~\ref{gap_vs_doping} also contains information on the angular dependence of the gaps in the Brillouin zone for the superconducting CDMFT solution. Fig.~\ref{gap_vs_doping}(d) shows that for every doping $p$, we can find a nodal gap (near ($\frac{\pi}{2},\frac{\pi}{2}$)), and an antinodal gap (near ($0,\pi$)), and that these gaps follow the same doping dependence as the sub-gap structure and the superconducting gap in Fig.~\ref{gap_vs_doping}(a).
Let us clarify the steps required to reach this conclusion. 
Fig.~\ref{gap_vs_doping}(e) defines the angle $\theta$ from the $(\pi,\pi)$ point of the Brillouin zone, and three straight paths (red, green, and yellow arrows), each along a different angle $\theta$. For each $\vec k$-point along these paths, Fig.~\ref{gap_vs_doping}(f) shows the $\omega$-dependent spectral weight. In these path-following spectral weights, we identify the smallest gap encountered (distance between the two maxima of the spectral weight closest to $\omega=0$), thus assigning one gap value for each angle~$\theta$. We can then plot the magnitude of the gap as a function of~$\theta$, as shown in Fig.~\ref{gap_vs_doping}(g) for $p=0.08$. As a function of angle, the gap starts at zero, because $\theta=0$ corresponds to the main $d$-wave node, then it grows, reaches a maximum at low~$\theta$ in the nodal region, and decreases back to zero, closing again at the extra nodes appearing in Figs.~\ref{bcs_cdmft}(d) and~\ref{sc_mdc}(b-c). The gap then grows back to reach its maximal value at large~$\theta$, in the anti-nodal region. This allows us to define the nodal gap as the local maximum of the gap between the main node and the extra node, and the anti-nodal gap as the maximal gap at large angle~$\theta$. In Fig.~\ref{gap_vs_doping}(d), we plot the magnitudes of these two gaps for all values of doping. They follow exactly the doping dependence of the sub-gap structure and the superconducting gap of Fig.~\ref{gap_vs_doping}(a). This leads us to associate the subgap structure to the nodal gap, and associate the full superconducting gap to the antinodal gap.
These associations are consistent with our interpretation of the subgap structure as saddle points in Fig.~\ref{sc_mdc}, because the latter correspond to the local maxima of the gap between the main node and the extra nodes.

After careful analysis of this section's results on the superconducting CDMFT solution, one should be convinced that the three main phenomena described up to here---the subgap structures, the extra nodes, and the nodal and antinodal gaps---are all different facets of how the spectral weight is shaped by the cluster density waves. They are consequences of CDMFT artificially breaking translation invariance.

\subsection{Fortuitous Agreement With STM Experiments}\label{experiments}\label{sec_discussion}

With this full picture of how CDMFT breaks translation invariance, we can now describe its implications for the agreement between CDMFT and some experimental observations on cuprates. We focus on two important results: the nodal-anti-nodal dichotomy of the superconducting gap seen in ARPES, and the vortex subgap states seen in STM, briefly reviewed below.
Of course, since the density waves causing these phenomena in CDMFT are artificial, this comparison to experiments requires a nuanced discussion. Let us first review what is seen in experiments.

The experimental nodal-anti-nodal dichotomy of the gap is related to the famous Fermi arcs observed in ARPES. Above $T_c$ (in the normal state), these arcs consist of an incomplete Fermi surface, taking the shape of four arcs~\cite{marshall_unconventional_1996, norman_destruction_1998}. These arcs appear in the nodal regions of the Brillouin zone (centered around $\vec k=(\pm\frac{\pi}{2},\pm\frac{\pi}{2})$), while a gap (the pseudogap) persists high above $T_c$ in the anti-nodal regions (around $\vec k=(\pm\pi,0)$ and $\vec k=(0,\pm\pi)$)~\cite{reber_origin_2012}. Although this dichotomy between the nodal gap and the anti-nodal gap is observed above $T_c$, it also persists below $T_c$, in the superconducting state. Indeed, in under-doped samples, the $d$-wave superconducting gap is not perfectly $d$-wave: it overshoots in the anti-nodal region~\cite{yoshida_pseudogap_2011, hashimoto_energy_2014}. This is what is meant by ``nodal-anti-nodal dichotomy'' of the superconducting gap: an imperfect $d$-wave gap with two separate magnitudes; smaller than expected near the node, larger than expected near the antinode.

The experimental subgap structures are observed in the gap measured by STM. The subgap structures are sometimes called ``vortex states'' or ``vortex excitations'' for reasons that will become clear below. The gap seen by STM is highly inhomogeneous: it differs as much from site to site on the lattice than the average gap differs from doping to doping~\cite{mcelroy_coincidence_2005}. Among the various spectra observed from site to site, one spectrum is preeminently associated to the pseudogap. This pseudogap spectrum is found when superconductivity is weak, for example in the vortex cores of magnetic fields~\cite{pan_stm_2000, hoogenboom_low-energy_2000, hoffman_four_2002, levy_fourfold_2005}, at positions where charge order is dominant~\cite{hoffman_four_2002, levy_fourfold_2005, kohsaka_intrinsic_2007}, or simply in very underdoped samples~\cite{mcelroy_coincidence_2005}. This low-temperature pseudogap spectrum shows two gaps: a large gap with rounded coherence peaks, and a smaller gap developing inside the large one. The smaller gap is the one referred to as the subgap structures of the superconducting gap.

\begin{figure}
\includegraphics[width=\columnwidth]{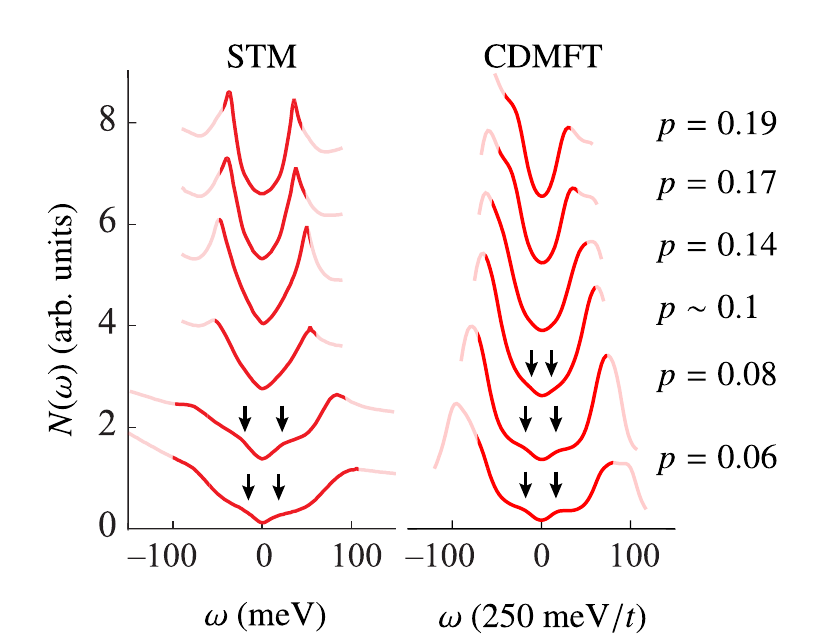}
\caption{
Fortuitous agreement with 
low temperature STM gaps at various dopings (data from Ref.~\cite{kohsaka_how_2008}) showing low-energy subgap structures similar to those found in the gaps of our superconducting CDMFT solutions at zero temperature.
}
\label{gap_in_brillouin}
\end{figure} 

To summarize experimental results, both ARPES and STM experiments find double gaps at low temperature in the superconducting state of cuprates: ARPES observes a nodal-anti-nodal dichotomy and STM observes subgap structures. We already discussed extensively our equivalent findings in CDMFT: the previous section demonstrated the presence of subgap structures, and of nodal and anti-nodal gaps in the CDMFT superconducting solution. Furthermore,
as we described, the key element causing subgap structures and nodal-anti-nodal gaps in CDMFT are the artificial $2\times 2$ density waves intrinsic to our method. This strongly suggests that the key element causing the experimentally observed subgap structures, and nodal-anti-nodal gaps in STM and ARPES, might be the $4\times 4$ charge order experimentally observed in cuprates, as reviewed in our introduction~\cite{comin_resonant_2016, hoffman_four_2002,kohsaka_intrinsic_2007,ghiringhelli_long-range_2012,hamidian_atomic-scale_2016,comin_broken_2015,comin_resonant_2016,fujita_direct_2014,comin_symmetry_2015,hamidian_atomic-scale_2016}.


To illustrate this compelling comparison of CDMFT to experiments, Fig.~\ref{gap_in_brillouin} presents our CDMFT density of states next to STM conductance curves at various dopings. Both show subgap structures with similar doping dependences.
It is not clear why the $2\times 2$ periodicity of CDMFT produces similar subgap structure as those seen in experiments, given the different periodicity of approximately $4\times 4$ in the latter.
Indeed, the mean-field models of Ref.~\cite{verret_subgap_2017} showed that usually, substructures caused by density waves are extremely sensitive to their periodicity. Subgap structures being so similar in CDMFT and experiments may be an indication that once translation invariance is broken, the interaction causes a certain balance in the density of states as a function of doping, regardless of the periodicity.

\begin{figure}
\includegraphics[width=\columnwidth]{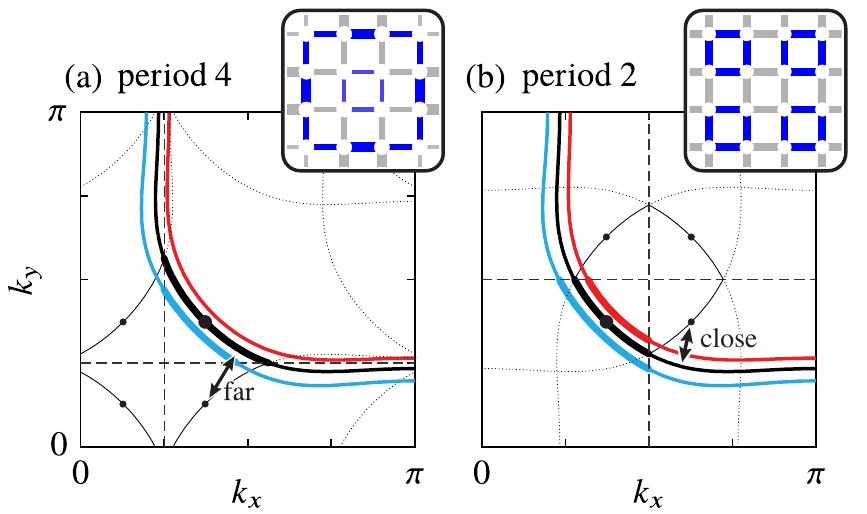}
\caption{
Position of the first Bragg planes (dashed lines) relative to the Fermi surface for half-filling (red), $p=0.125$ (black), and $p=0.25$ (blue). For doping $p=0.125$ (black), the position of possible extra nodes is illustrated on the corresponding copies of the Fermi surface (thin lines and dotted line). These copies are given by $t(\vec k + \vec Q_x)$, $t(\vec k + \vec Q_y)$, and $t(\vec k + \vec Q_x + \vec Q_y)$ (a) for $\vec Q_x=(1/4,0)2\pi$ and $\vec Q_y=(0,1/4)2\pi$, as in cuprates (the inset shows a representation of the bidirectional $d$-form-factor density waves) and (b) for $\vec Q_x=(1/2,0)2\pi$ and $\vec Q_y=(0,1/2)2\pi$, as in CDMFT (the inset shows the $2\times2$ cluster tilling).
}
\label{q4vsq2}
\end{figure} 

Despite this agreement for subgap structures, the nodal-antinodal dichotomy seen in ARPES differs a lot from the one obtained in our CDMFT results. The most notable difference is the presence of the aforementioned extra nodes, which are obtained in CDMFT, but are not seen in experiments. We have three possible explanations for this, all related to differences between the artificial $2\times2$ density waves present in CDMFT and the $\sim4\times4$ charge order observed experimentally. First, as illustrated in Fig.~\ref{q4vsq2}, the position of eventual extra nodes in the $4\times 4$ case would be farther from the bare Fermi surface than in the $2\times 2$ case, especially at low dopings. This could lead to negligible spectral weight at these extra nodes in the experiments. 
Second, the breaking of translation invariance in CDMFT is rather strong as indicated by the high mean-field gaps (Weiss fields) required to reproduce the CDMFT results: 0.2 and 0.125 for charge and pair density waves compared to 0.07 for superconductivity. By contrast, in cuprates the characteristic energy of charge order is believed to be of the same order as that of superconductivity~\cite{chang_direct_2012}, which might lead to weaker reconstruction of the spectral weight.
Third, the experimental density waves are short-ranged and might therefore also lead to weaker reconstruction of the spectral weight. These considerations might explain why features as sharp as the extra nodes found here are not seen in ARPES. 

Let us also compare how the Bragg planes intersect the Fermi surface in Fig.~\ref{q4vsq2} as a function of doping for the $2\times2$ case and the $4\times4$ case. The lengths of Fermi surface between the Bragg planes follow opposite dependencies: in the $4\times 4$ case, this length decreases with doping, whereas in the $2\times 2$ case it increases with doping. This difference may be relevant when studying Fermi arcs with these methods.

We must also mention that a nodal-antinodal dichotomy similar to that found in our CDMFT results had already been studied by Aichhorn et al. using variational cluster approximation (VCA) with a $3\times3$ cluster~\cite{aichhorn_superconducting_2007}. Their paper did not consider broken translation invariance as a cause for the dichotomy. Yet, in the first figure of their paper, the Bragg planes of the $3\times3$ periodicity can clearly be identified and, revisiting the paper with our results in mind, it is possible that the broken translation invariance is entirely responsible for the reported dichotomy of the superconducting state. Moreover, as expected from our discussion, their results with a $3\times3$ cluster agrees better with experiments than our $2\times2$ results, arguably because it is closer to the $4\times4$ experimental case.


\section{Conclusion}\label{conclusion}
In this paper, we explained the origin of subgap structures in the superconducting gap of CDMFT solutions to the Hubbard model. These subgap structures are a consequence of broken translation invariance due to the self-energy being zero between clusters in CDMFT. To quantify this broken translation invariance, we measured the order parameter of $2\times2$ charge density waves in both the normal and the superconducting CDMFT solutions, along with $2\times2$ pair density waves in the superconducting case. These order parameters are weaker at higher doping, as determined by the amplitude of the self-energy which decreases with doping. Careful examination of the spectral weight in the superconducting CDMFT solution showed that the subgap structures appear because of new Bragg planes introduced by the superlattice used in CDMFT. We also showed that, at low energy, the results for the spectral weight at small $\eta$ can be reproduced with a mean-field model of coexisting superconductivity and $2\times2$ density waves, leading to the same subgap structures as in CDMFT. We also showed that varying the Lorentzian broadening $\eta$ is a powerful tool to discriminate artefacts from robust features. 

Although the presence of these artificial density waves is an uncontrolled by-product of the CDMFT approximation, studying its interplay with superconductivity is enlightening. We discussed subgap structures observed in STM experiments on cuprates that are strikingly similar to those found in our CDMFT results. We also showed that the subgap structures found in CDMFT are related to a dichotomy between nodal and anti-nodal superconducting gaps, reminiscent of the one observed in ARPES, albeit much sharper in CDMFT. Our results therefore suggest that these ``double gaps'' observed experimentally by STM and ARPES could be indirect consequences of the coexistence between density waves and superconductivity in cuprates (both due to strong correlations).

To conclude, let us clarify this last point. We believe this is how our work allows to disentangle the genuine strongly correlated state captured by CDMFT (pseudogap and $d$-wave superconductivity), from indirect effects coming form broken translation invariance (here subgap structure and nodal-antinodal dichotomy in the superconducting state). As we said in the introduction, pseudogap physics at a high temperature $T^*$ is mostly captured by strongly correlated electron methods (CDMFT and DCA) without the need for spontaneously broken symmetry. Although we did not discuss DCA extensively, we explained that it preserves translation invariance. CDMFT, on the other hand, artificially breaks translation invariance, which was the subject of this paper. The pseudogap, as manifested in the drop of the Knight shift at high temperature $T^*$, is qualitatively consistent in these two methods, which suggests that its fundamental physics is independent of whether translation invariance is broken or not. However, the two low-temperature experimentally observed phenomena we reviewed---subgap structures and nodal-anti-nodal dichotomy of the superconducting gap---were obtained in zero-temperature CDMFT solely because of broken translation invariance, as demonstrated throughout this paper. These two phenomena are often considered landmarks of pseudogap physics in cuprates. Our results suggest that they are distinct from the strongly correlated pseudogap captured in CDMFT and DCA at $T^*$. They are instead consequences of inhomogeneity (charge or pair order) at low temperature $T\rightarrow 0$ that modify the superconducting gap. To confirm this hypothesis, we would need a proof that subgap structures and nodal-antinodal dichotomy of the superconducting gap do not appear in a translation-invariant low-temperature pseudogap (if such a thing exists). Unfortunately, this last point requires a strongly correlated theory capturing the pseudogap while preserving translation invariance \emph{and} providing sufficient $\vec k$-resolution of the self-energy. Such a method does not exist at the moment, and thus new developments in this direction are still needed.



\begin{acknowledgments}
We acknowledge Peayush Choubey, Seamus Davis, Stephen Edkins, Patrick Fournier, Antoine Georges, Gaël Grissonanche, Jan Guckelberger, Mohammad Hamidian, Charles-David Hébert, Jennifer Hoffman, Marc-Henri Julien, Gabriel Kotliar, David LeBoeuf, Reza Nourafkan, Catherine Pépin, Olivier Simard, Louis Taillefer, and Tatiana Webb for fruitful discussions.
This research was undertaken thanks in part to funding from the Canada First Research
Excellence Fund, the Natural Sciences and Engineering Research Council (Canada) under grants RGPIN-2014-04584, RGPIN-2015-05598 and RGPIN-2019-05312, and the Fonds Nature et Technologie (Qu\'ebec). Computing resources were provided by Compute Canada and Calcul Qu\'ebec
\end{acknowledgments}

\appendix

\section{Signatures of Broken Translation Invariance in the Periodized Cumulant and Self-Energy}\label{a1}

The results of section \ref{section_periodized} demonstrate that the periodized Green's function $G(\vec k,z)$ still contains signatures of broken translation invariance. In this appendix, we want to show that such signatures also exist for the periodized cumulant $M(\vec k,z)$ and self-energy $\Sigma(\vec k,z)$.

In contrast to $G(\vec k,z)$, which is obtained from periodizing the already $\tilde{\vec k}$-dependent $\vec G(\tilde{\vec k},z)$, the periodized cumulant $M(\vec k,z)$ and self-energy $\Sigma(\vec k,z)$ are obtained by periodizing $\vec M(z) = (z + \mu - \vec\Sigma^{\text{c}}(z))^{-1}$ and $\vec\Sigma^{\text{c}}(z)$ which are both constant in $\tilde{\vec k}$.

When periodizing a quantity constant in $\tilde{\vec k}$, the signatures of broken translation invariance are only constant factors modifying Fourier components. Let us illustrate what this means with the example of the cluster hopping~$\vec t^{c}$. Like $\vec M(z)$ and $\vec \Sigma^{c}(z)$, the hopping $\vec t^{c}$ is constant in $\tilde{\vec k}$ and breaks translation invariance by being zero between clusters. For our $2\times 2$ cluster ($N_c=4$), periodizing $\vec t^{c}$ yields
\begin{align}
t^{\text{per}}(\vec k) 
&= 
\frac{1}{N_c}\sum_{\vec R,\vec R'}
e^{-i\vec k\cdot(\vec R-\vec R')}
t^{c}_{\vec R,\vec R'}
\\
&=
-t(\cos k_x+\cos k_y) - t'\cos k_x\cos k_y.
\label{false} 
\end{align}
On the other hand, if we periodize the lattice hopping $\vec t(\tilde{\vec k})$,
\begin{align}
t(\vec k) 
&= 
\frac{1}{N_c}\sum_{\vec R,\vec R'}
e^{-i\vec k\cdot(\vec R-\vec R')}
t_{\vec R,\vec R'}(\tilde{\vec k})
\\
\nonumber &=
 -2t(\cos k_x+\cos k_y) - 4t'\cos k_x\cos k_y\\
&\qquad-2t''(\cos k_x + \cos k_y)+\cdots
\label{true}
\end{align}
we recover the true dispersion. As one can see, the difference between~\eqref{false} and~\eqref{true} are prefactors (with prefactor zero for $t''$ and beyond since they are absent from the cluster).
These incorrect prefactors in~\eqref{false} show that the periodization of a non-translation-invariant quantity does not yield the correct result.
The analog of these prefactors can also be expected in $M(\vec k, z)$ and $\Sigma(\vec k, z)$. This is what we mean by signatures of broken translation invariance in these cases.

Unfortunately, because of the inversions entering the definition of the Green's functions in~\eqref{sigma_periodization} and~\eqref{cumulant_periodization}, even such simple signatures can have drastic consequences. For example, periodizing the self-energy leads to the loss of the Mott gap in one dimension~\cite{senechal_cluster_2012-1}. Periodizing the cumulant can also lead to unphysical results~\cite{simard_master}. Moreover, trying to correct these factors by replacing the periodization by an expression inspired of~\eqref{true} was attempted for the self-energy~\cite{lichtenstein_antiferromagnetism_2000}, but later shown to violate causality~\cite{biroli_cluster_2002}.

\bibliography{verret_intrinsic_2017}

\end{document}